\documentclass[lettersize,journal]{IEEEtran}
\IEEEoverridecommandlockouts
\usepackage{float}
\usepackage{amsmath,amsfonts}
\usepackage{algorithmic}
\usepackage{algorithm}
\usepackage{array}
\usepackage[caption=false,font=normalsize,labelfont=sf,textfont=sf]{subfig}
\usepackage{stfloats}
\usepackage{url}
\usepackage{cite}
\usepackage{amsmath,amssymb,amsfonts}
\usepackage[hidelinks,urlcolor=blue]{hyperref} 
\usepackage{graphicx}
\usepackage{textcomp}
\usepackage{xcolor}
\usepackage{gensymb}
\usepackage{algorithm}
\usepackage{algorithmic}
\graphicspath{{figures/}}
\def\BibTeX{{\rm B\kern-.05em{\sc i\kern-.025em b}\kern-.08em
    T\kern-.1667em\lower.7ex\hbox{E}\kern-.125emX}}

\begin{document}

\title{Modeling of Dark Count Probability in Perimeter-Gated SPADs}

\author{Md Sakibur Sajal,~\IEEEmembership{Student Member,~IEEE}, Ziyad Alswaidan,~\IEEEmembership{Student Member}, Tathagata Srimani,~\IEEEmembership{Senior Member,~IEEE}, and Marc Dandin,~\IEEEmembership{Senior Member,~IEEE}
\thanks{The authors are with the Department of Electrical and Computer Engineering, Carnegie Mellon University,  Pittsburgh, PA 15213 USA (e-mail: mdandin@andrew.cmu.edu). This material is based on work supported by the National Science Foundation
under Grant No. 2442346. Any opinions, findings, and conclusions or recommendations expressed in this material are those of the authors and do not necessarily reflect the views of the National Science Foundation.}
}

\maketitle

\begin{abstract}
This Letter presents a novel analytical framework showing that the dark count probability ($\boldsymbol{P}_{\boldsymbol{DC}}$) of perimeter-gated single-photon avalanche diodes (pg-SPADs) follows a complementary Gompertz function. Specifically, we show that $\boldsymbol{P}_{\boldsymbol{DC}}$ follows a complementary Gompertz form from which we derive a pixel-specific descriptor, the midpoint perimeter gate voltage, which characterizes  a pixel's equiprobable operating point.  We further show that a perimeter gate voltage compensation rate may be obtained from this descriptor to offset temperature-induced changes in the pixel's activation function. The proposed framework is experimentally validated using 4,096 pg-SPADs arranged in a $\boldsymbol{64 \times 64}$ array and manufactured in a 0.35~$\boldsymbol \mu$m CMOS process. The devices were characterized at temperatures ranging from $\boldsymbol {-5\,^{\circ}\mathrm{C}}$ to $\boldsymbol{ 55\,^{\circ}\mathrm{C}}$ and perimeter gate voltage magnitudes of 0 to 5~V. The measured results demonstrate deterministic bias control of dark count probability across process and temperature variations.
\end{abstract}


\section{Introduction}

Single-photon avalanche diodes (SPADs) enable photodetection with high temporal resolution but are inherently susceptible to noise originating from spurious avalanche events not directly caused by photon absorption\cite{s23073412,11107005}. Perimeter-gated SPADs (pg-SPADs) are structurally similar to conventional SPADs, with the only difference being a polysilicon gate disposed at the periphery of the planar avalanche photodiode~\cite{Dandin2010,Dandin2016,6165634md}. Therefore, fundamentally, pg-SPADs are prone to the same type of noise. 

However, unlike classical SPADs, where avalanche initiation is governed primarily by the applied reverse bias, the perimeter gate in pg-SPADs provides an additional control dimension for directly modulating the probability of avalanche. This implies that, \textit{ceteris paribus},  a pg-SPAD's dark count rate (DCR), \textit{i.e.}, its noise, may be controlled through perimeter gating~\cite{Sajal2022Perimeter-GatedProbability}.

\begin{figure}[!ht]
     \centering
    \includegraphics[width= 1\linewidth]{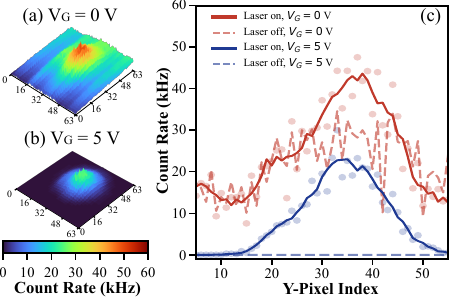}
\caption{Spatial and one-dimensional characterization of a perimeter-gated SPAD imager operated at extreme gate biases.
(a) Count rate surface at $V_G = 0$, showing a dark count-dominated response. 
(b) Count rate surface at $V_G = 5$, demonstrating near-complete suppression of dark counts while preserving photon sensitivity. A common color scale is used for (a) and (b).
(c) One-dimensional slice (row $x=32$) comparing laser on and laser off (dark) conditions for $V_G = 0$ and $V_G = 5$, illustrating strong suppression of dark counts and enhanced contrast at high gate bias. 
}
   
    \label{fig:fig1}
      \vspace{-18pt}
\end{figure}

This feature is illustrated experimentally in Figures~\ref{fig:fig1}(a),(b) with measured results from a $64\times64$ pg-SPAD array fabricated in a $0.35\,\mu$m CMOS process. As the perimeter gate voltage magnitude $V_G$ increases from $0$ to $5\,\mathrm{V}$, the measured count rate surfaces show a suppression of dark count activity across the array while maintaining a localized response to a focused laser spotlight. In the high-gating regime, the devices exhibit nearly vanishing dark count activity, yet photon-induced avalanche events remain detectable. These measurements indicate that the perimeter gate effectively modulates the stochastic triggering of avalanches without eliminating photon sensitivity.

The line profiles in Figure~\ref{fig:fig1}(c) further reveal an important consequence of perimeter gating: the signal-to-noise ratio (SNR) of the device becomes tunable through the gate bias. At higher $V_G$, the overall count rate decreases as dark counts are suppressed, while the laser-induced signal remains observable above the background. For $V_G = 5 ~V$, the dark count rate baseline approaches zero, whereas a distinct signal peak appears where the laser spotlight is present. This indicates that pg-SPADs enable adaptive control of detection operating points, allowing the device to trade sensitivity and noise through perimeter gate biasing.

Designing such adaptive operation requires a quantitative understanding of how dark count activity varies with gate voltage and how the underlying avalanche probability transitions between inactive and active regimes. A substantial body of literature has examined the statistics of dark counts in conventional SPADs~\cite{pdc2003,pdcgated,Zavvari:13,pdcref1,pdcref2}. In particular, these prior studies have modeled dark count probability in terms of the underlying dark carrier generation mechanisms and the probability that a generated carrier initiates an avalanche event.

However, very little work has been done on modeling the gate voltage dependence of the dark count rate \textit{probability} of pg-SPAD devices~\cite{sajalpbit,spadpuf}. This Letter bridges this gap by presenting a new analytical framework that accounts for the effects of perimeter gating on pg-SPAD DCR probability.

\begin{figure}[!t]
    \centering
    \includegraphics[width=\linewidth]{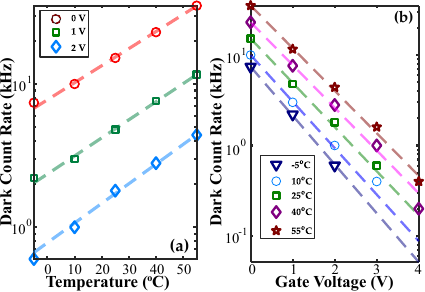}
   \caption{Measured perimeter gating and dark count rate (DCR) dynamics of a pg-SPAD from the array. (a) DCR increased exponentially with temperature for fixed perimeter gate voltages. (b) DCR decreased exponentially with increasing perimeter gate voltage magnitude for fixed operating temperatures. Dashed lines indicate exponential fits to the measured data. (Missing data markers indicate that no count was registered during measurement.)}
    \label{fig:DCR_DYNAMICS}
\end{figure}

We develop a model that can be parameterized to account for the dispersion in dark count probability typically observed across large pg-SPAD arrays due to manufacturing process variations. We also show that the model captures the dependence of dark count probability on temperature, providing a means for dynamic compensation of DCR probability under temperature fluctuations. The proposed framework is experimentally validated using a $64\times64$ pg-SPAD array with monolithically integrated peripheral readout circuitry, fabricated in a commercial $0.35~\mu$m CMOS process.

\section{Dark Count Probability Model}
\label{sec:2}

 \subsection{Empirical DCR Model}
We first develop an empirical pg-SPAD DCR model that depends on the perimeter gate voltage and temperature, taking as a starting point the measured data shown in Figure~\ref{fig:DCR_DYNAMICS}. In particular, the data show that, similar to conventional SPADs~\cite{stdSpadDcr}, the average pg-SPAD DCR, $\lambda$,  increases exponentially with temperature. Furthermore, the data show that the gate voltage has a suppressive effect despite the DCR's strong temperature dependence~\cite{10375697}. 

The model thus includes two components, one that captures the DCR's perimeter gate voltage dependence and another that captures its temperature dependence. Specifically, Equation~\ref{onlyC} captures the dependence of the DCR on temperature (Figure~\ref{fig:DCR_DYNAMICS}(a)). And, Equation~\ref{onlyVg} brings the dependence on the perimeter gate voltage to the fore (Figure~\ref{fig:DCR_DYNAMICS}(b)). For each component, we varied temperature and gate voltage to capture the combined effect of both parameters. 

Combining Equations~\ref{onlyC}~and~\ref{onlyVg}, we obtain Equation~\ref{eq:whole}, which is a unified model of the DCR dynamics. The parameters $\alpha$ and $\zeta$ are fitting constants, $\theta$ is the temperature, and $\lambda_{0,0}$ is chosen as the DCR at a gate voltage of $V_G = 0~V$  and at $0~^{\circ}C$. From the measured data for all of the pg-SPAD devices in the array, we observed that $\lambda_{0,0}$, $\zeta$, and $\alpha$  vary from device to device due to process variations, as we shall discuss in detail~below.

\begin{eqnarray}
    \lambda(\theta,V_G) &=& \lambda_{0} (V_G)e^{\zeta\theta}
    \label{onlyC} \\
    \lambda_0(V_G) &=& \lambda_{0,0}e^{-\alpha V_G}
    \label{onlyVg} \\
    \lambda(\theta,V_G) &=& \lambda_{0,0}e^{\zeta\theta-\alpha V_G}
    \label{eq:whole}
\end{eqnarray}

\subsection{Gompertz Model of Dark Count Probability}

We now make use of the fact that avalanche events in Geiger mode devices follow Poisson statistics~\cite{Poisson}. This means that the probability of registering $X = k$ events is given by the Poisson probability mass function shown in Equation~\ref{eq:poisson_pmf_time}, where $T$ is the integration time\footnote{Here, we borrow the term ``integration time'' from active pixel sensor imagers to indicate a pixel exposure time; however, it is noted that a pg-SPAD front-end does not integrate charge, rather, it produces a digital signal whose leading edges correspond to avalanche events.}, and $\lambda$ is the expected dark count rate (as denoted above in our empirical DCR model).  Therefore, because the probability of having no events ($k= 0$) is $e^{-\lambda T}$, we define the probability of having at least one dark count, $P_{DC}$, as the complementary probability of having no events. This is shown in Equation~\ref{eq:pdc_main}.

\begin{equation}
P(X=k)=\frac{(\lambda T)^k e^{-\lambda T}}{k!},
\qquad k=0,1,2,\ldots
\label{eq:poisson_pmf_time}
\end{equation}

\begin{equation}
    P_{DC} = 1 - e^{-\lambda T}
    \label{eq:pdc_main}
\end{equation}

Substituting Equation~\ref{eq:whole} in Equation~\ref{eq:pdc_main}, we obtain: 

\begin{equation}
    P_{DC} = 1 - e^{-\bar{\beta} e^{\zeta\theta-\alpha V_G}}
    \label{eq:pdc_expand}
\end{equation}
where $\bar{\beta} = \lambda_{0,0} T$. Further, for a fixed temperature, we write:
\begin{equation}
    P_{DC} = 1 - e^{-{\beta} e^{-\alpha V_G}},
    \label{eq:pdc_gomp}
\end{equation}
where $\beta = \bar{\beta} e^{\zeta\theta}$.

Equation~\ref{eq:pdc_gomp} follows the form of a typical complementary Gompertz function~\cite{Winsor1932TheCurve,sajalpbit}. This result implies that a pg-SPAD's dark count probability follows a sigmoidal activation function; and, specifically,  for fixed temperature and integration time, this probability can be \textit{tuned} using the perimeter gate voltage.

\subsection{Dark Count Probability Temperature Compensation}
From the Gompertz relation shown in Equation~\ref{eq:pdc_gomp}, we define a pixel-specific physical descriptor, the midpoint  voltage, $V_{G,mid}$, which is the perimeter gate voltage at which $P_{DC} = 0.5$; at that perimeter gate voltage, it becomes equi-probable to register a dark count or no dark count. Equation~\ref{eq:midpoint} is obtained by setting Equation~\ref{eq:pdc_gomp} equal to 0.5.

\begin{equation}
    V_{G,mid} = \frac{ln(k\beta)}{\alpha}, k = \frac{1}{ln(2)}
    \label{eq:midpoint}
\end{equation}

By substituting $\beta = \lambda_{0,0}Te^{\zeta\theta}$ in Equation~\ref{eq:midpoint}, we obtain the functional relationship shown in Equation~\ref{eq:vmidexpand}, which relates $V_{G,mid}$ to other model parameters.

\begin{equation}
    V_{G,mid} = \frac{ln(k) + ln(\lambda_{0,0}) + ln(T) + \zeta\theta} {\alpha}.
    \label{eq:vmidexpand}
\end{equation}

This result suggests that $V_{G,mid}$ depends  on the logarithm of the integration time $T$ and that it changes with temperature $\theta$ by a factor of $\frac{\zeta}{\alpha}$. This factor, which we call the temperature correction coefficient $\gamma$, denotes the adjustment needed to maintain the same $V_{G,mid}$ when the temperature $\theta$ changes. In other words, it is the rate of the change of $V_{G,mid}$ with respect to temperature (Equation~\ref{eq:compensation}), and it can be used to offset changes in temperature. 

\begin{equation}
    \gamma = \frac{d}{d\theta}(V_{G,mid}) = \frac{\zeta}{\alpha}
    \label{eq:compensation}
\end{equation}

\section{Experimental Validation}

In order to validate the pg-SPAD dark count probability Gompertz model shown in Equation~\ref{eq:pdc_gomp} and its corollary results shown in Equations~\ref{eq:midpoint}-\ref{eq:compensation}, we devised an experimental approach for measuring $P_{DC}$. From experimentation, we set $P_{DC}$ as a the sum of a Boolean variable $m$  to the total number of observations $N$, each observation having the same integration time $T$, and where $m$ is a Boolean variable that indicates whether at least one dark event was registered (Equation~\ref{eq:pdc_sum}).

\begin{figure}[!t]
    \centering
    \includegraphics[width=\linewidth]{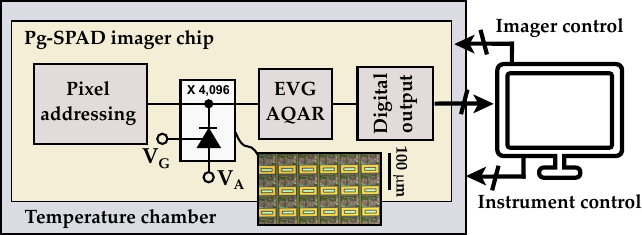}
    \vspace{-15pt}
   \caption{Experimental setup and simplified system-level block diagram of the pg-SPAD imager. The $64\times64$ pg-SPAD array was operated inside a temperature chamber while the perimeter gate voltage $V_G$ and reverse bias voltage $V_A$ were externally controlled. Pixel addressing, event generation (EVG), active quench and active reset (AQAR), and digital readout circuitry were monolithically integrated on chip. Data acquisition, instrument control, and processing were performed through a computer-controlled measurement platform. The inset shows a photomicrograph of a set of pixels from the fabricated chip.}
    \vspace{-10pt}
    \label{fig:blockDiagram}
    \end{figure}

\begin{equation}
P_{DC} = \frac{1}{N}\sum_{i=1}^{N} m_i ,
\label{eq:pdc_sum}
\end{equation}

Here, we set $m_i = 1$ if there was at least one dark event detected during the $i$-th observation and to $0$ otherwise. We reiterate that all observations are made under a fixed set of operating parameters (\textit{i.e.}, fixed bias voltages, temperature, and integration time). Measurements were conducted in a dark temperature chamber (EC1X, Sun Electronic Systems, Inc.), as shown in Figure~\ref{fig:blockDiagram}.

\begin{figure}[!t]
    \centering
    \includegraphics[width=1\linewidth]{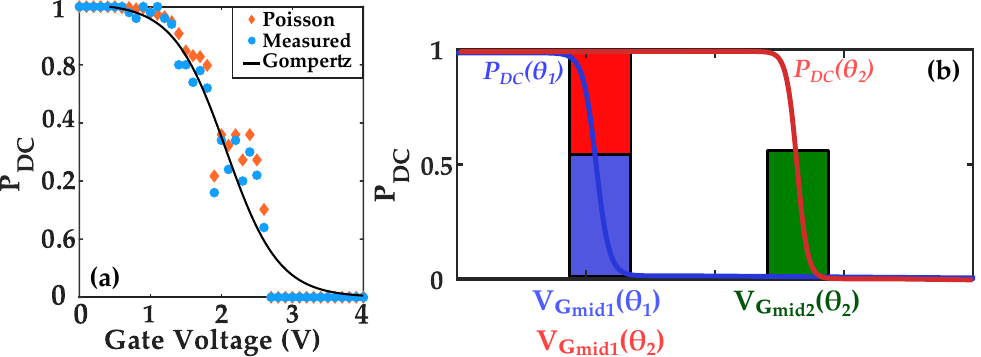}
    \caption{Experimental validation of the Gompertz model and  of the proposed temperature compensation scheme. (a) Activation function of a pg-SPAD chosen at random, comparing the Gompertz model, the Poisson estimates, and the experimentally measured $P_{DC}$. (b) Adjustment of the perimeter gate voltage to achieve equiprobable dark count probability in view of a temperature shift from $\theta_1$ to $\theta_2$, where $\theta_2 = 35~^\circ C$ and  $\theta_1= 25~^\circ C$.}
    \label{fig:gompComp}
\end{figure}

Figure~\ref{fig:gompComp} (a) shows the experimental validation of the Gompertz dark count probability model formulated in Equation~\ref{eq:pdc_gomp}. The results are for a pg-SPAD chosen at random in the array, for an integration time of $T = 5~ms$ and at room temperature ($25~^\circ C$). These values correspond to $\alpha = 1.4~V^{-1}$ and $\beta = 11.8$ in the Gompertz model. 

The data show that the experimentally measured $P_{DC}$ generally agrees with the Poisson statistics. Furthermore, the data show that the Gompertz model also tracks the experimentally-measured values as well as the Poisson estimates. We verified this for each pixel in the array, \textit{i.e.}, an appropriate set of fitting constants could be found to track the experimentally-derived $P_{DC}$ as well as the Poisson estimates for each of the 4,096 pixels.  

Moreover, as expected, varying the temperature shifted the $P_{DC}$ vs. gate voltage curve to the right, with increasing temperature.  This means that for a given $V_{G,mid}$ at a lower temperature, the temperature  correction coefficient $\gamma$ can be used to obtain a new $V_{G,mid}$ for achieving equiprobable dark count registration at the elevated temperature. This is illustrated in Figure~\ref{fig:gompComp}(b) and is described in further detail below in Equation~\ref{eq:Correction}.

\begin{equation}
    V_{G,mid_{2,| \theta_2}} = V_{G,mid_{1,| \theta_1}}  + \gamma \delta \theta,~   \delta \theta = \theta_2 - \theta_1
    \label{eq:Correction}
\end{equation}

\begin{figure}[!ht]
    \centering
    \includegraphics[width=1\linewidth]{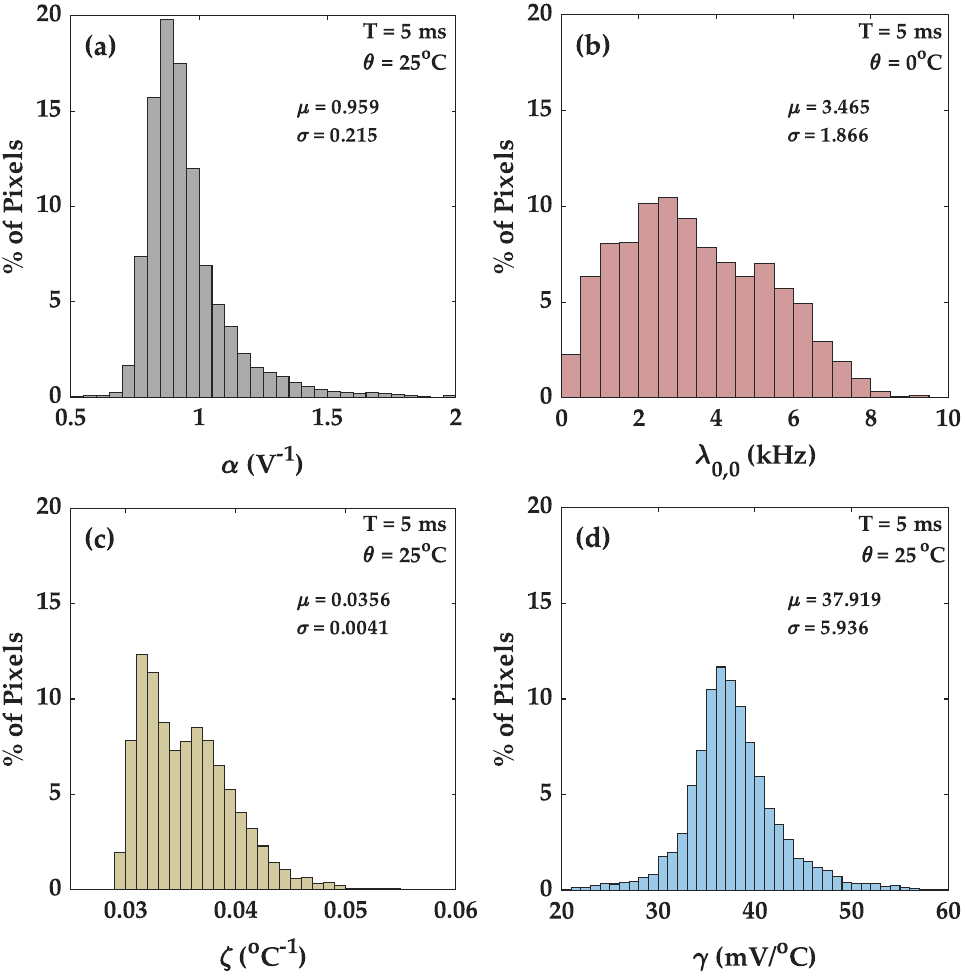}
    \caption{Experimental characterization of model parameter dispersion due to process variations. (a)-(c) Gompertz model fitting constants for fixed temperature and integration time. The expected dark count rate $\lambda_{0,0}$ is measured at $0^{\circ} C$. (d) Measured variation in the temperature correction coefficient $\gamma$ across the pixel array.} 
    \label{fig:modelVariations}
\end{figure}

Turning now to Figure~\ref{fig:modelVariations}, we discuss the dispersion in the model parameters. Specifically, we show that there is a distribution of values for each of $\alpha$, $\zeta$, $\lambda_{0,0}$. The same is also true for the temperature correction coefficient $\gamma$. We posit that these distributions arise from the non-idealities of the CMOS manufacturing process. In other words, there is an inherent mismatch between the pixels in the array, and as such, each pixel exhibits a specific activation function. Therefore, it follows that each pixel has a specific Gompertz model.

\section{Conclusion}

This Letter presented a compact analytical framework for modeling dark count activation in perimeter-gated SPAD imagers. By combining the measured dependence of DCR on perimeter gate voltage and temperature with Poisson event statistics, we showed that the dark count probability $P_{DC}$ follows a complementary Gompertz function. This result provides a simple analytical interpretation of perimeter gating: the gate does not merely suppress dark counts, but continuously tunes the stochastic activation function of each pg-SPAD. The resulting midpoint voltage $V_{G,\mathrm{mid}}$ defines a pixel-specific equiprobable operating point, while the temperature correction coefficient $\gamma=\zeta/\alpha$ provides a direct bias adjustment for maintaining a prescribed dark count probability under temperature variation.

The model was experimentally validated using a $64\times64$ pg-SPAD imager fabricated in a $0.35\,\mu$m CMOS process. The measured data showed that the Gompertz formulation captures the dark count probability of individual pixels and that the extracted parameters vary across the array due to process-induced mismatch. These parameter distributions provide a practical basis for calibration, compensation, and first-order statistical simulation of large pg-SPAD imagers.

Beyond photon detection, the framework is relevant to emerging pg-SPAD applications in which dark counts are intentionally used as stochastic events rather than treated only as noise~\cite{spadpuf,sajalpbit,10375697}. In probabilistic bits~\cite{sajalpbit} and true random number generators~\cite{10375697}, useful operation depends on the ability to set, stabilize, and reproduce an output probability or entropy source across devices and operating conditions. The proposed model provides this missing link by mapping gate bias, temperature, and pixel-to-pixel variation to dark count probability.

Lastly, we caution that the extracted model parameters and their measured distributions should be interpreted as dependent on the characterization conditions under which they were obtained. For fixed temperature, integration time, and biasing conditions, each pixel can be described by a static complementary Gompertz activation function. However, the corresponding population distributions are expected to change with temperature, integration time, operating bias, and manufacturing conditions. Therefore, a complete statistical model for predictive array level simulation would require additional characterization across multiple chips, fabrication lots, and operating conditions.


\bibliographystyle{IEEEtranDOI}
\bibliography{main}

@ARTICLE{6165634md,
  author={Dandin, Marc and Abshire, Pamela},
  journal={IEEE Electron Device Letters}, 
  title={High Signal-to-Noise Ratio Avalanche Photodiodes With Perimeter Field Gate and Active Readout}, 
  year={2012},
  volume={33},
  number={4},
  pages={570-572},
  doi={10.1109/LED.2012.2186112}}

@INPROCEEDINGS{11107005,
  author={Calmon, Francis and Habach, Rida and Uhring, Wilfried and Pellegrini, Sara and Golanski, Dominique and Simiz, Jean-Gabriel and Mureddu, Ugo},
  booktitle={2025 23rd IEEE Interregional NEWCAS Conference (NEWCAS)}, 
  title={{Using the Dark Count Rate of SPAD Implemented in CMOS FD-SOI Technology for True Random Number Generator}}, 
  year={2025},
  volume={},
  number={},
  pages={485-489},
  doi={10.1109/NewCAS64648.2025.11107005}}

@ARTICLE{10375697,
  author={Sajal, Md. Sakibur and Dandin, Marc},
  journal={IEEE Transactions on Circuits and Systems II: Express Briefs}, 
  title={{True Random Number Generation Using Dark Noise Modulation of a Single-Photon Avalanche Diode}}, 
  year={2024},
  volume={71},
  number={3},
  pages={1586-1590},
  doi={10.1109/TCSII.2023.3347735}}

@Article{Poisson,
AUTHOR = {Zhang, Hanping and Zhu, Xinyi and Wang, Yurong and Wu, E and Wu, Guang},
TITLE = {{A Complementary Approach for Characterizing Dark Count Rate in First-Photon-Gated Single-Photon Detectors}},
JOURNAL = {Photonics},
VOLUME = {13},
YEAR = {2026},
NUMBER = {5},
ARTICLE-NUMBER = {468},
DOI = {10.3390/photonics13050468}
}

@Article{s23073412,
AUTHOR = {Qian, Xuanyu and Jiang, Wei and Elsharabasy, Ahmed and Deen, M. Jamal},
TITLE = {{Modeling for Single-Photon Avalanche Diodes: State-of-the-Art and Research Challenges}},
JOURNAL = {Sensors},
VOLUME = {23},
YEAR = {2023},
NUMBER = {7},
ARTICLE-NUMBER = {3412},
PubMedID = {37050472},
ISSN = {1424-8220},
DOI = {10.3390/s23073412}
}

@article{pdcref2,
  author    = {S. I. Soloviev and A. V. Vert and J. Fronheiser and P. M. Sandvik},
  title     = {{Solar-Blind 4H-SiC Avalanche Photodiodes}},
  journal   = {Materials Science Forum},
  volume    = {615-617},
  pages     = {873--876},
  year      = {2009},
  doi       = {10.4028/www.scientific.net/msf.615-617.873},
}

@article{pdcref1,
    author = {Comandar, L. C. and Fröhlich, B. and Dynes, J. F. and Sharpe, A. W. and Lucamarini, M. and Yuan, Z. L. and Penty, R. V. and Shields, A. J.},
    title = {{Gigahertz-Gated InGaAs/InP Single-Photon Detector with Detection Efficiency Exceeding 55$\%$ at 1550 nm}},
    journal = {Journal of Applied Physics},
    volume = {117},
    number = {8},
    pages = {083109},
    year = {2015},
    month = {02},
    doi = {10.1063/1.4913527}
}

@article{Zavvari:13,
author = {Mahdi Zavvari and Vahid Ahmadi},
journal = {Appl. Opt.},
number = {32},
pages = {7675--7681},
publisher = {Optica Publishing Group},
title = {{Quantum Dot Infrared Photodetector with Gated-Mode Design for Mid-IR Single Photon Detection}},
volume = {52},
month = {Nov},
year = {2013},
doi = {10.1364/AO.52.007675}
}

@ARTICLE{stdSpadDcr,
  author={Richardson, Justin A. and Grant, Lindsay A. and Henderson, Robert K.},
  journal={IEEE Photonics Technology Letters}, 
  title={{Low Dark Count Single-Photon Avalanche Diode Structure Compatible With Standard Nanometer Scale CMOS Technology}}, 
  year={2009},
  volume={21},
  number={14},
  pages={1020-1022},
  doi={10.1109/LPT.2009.2022059}}

@inproceedings{pdcgated,
author = {Alessandro Restelli and Joshua C. Bienfang and Alan L. Migdall},
title = {{Ultra-Short Gates Improve the Performance of High-Speed Gated Single-Photon Avalanche Diodes}},
volume = {9114},
booktitle = {Advanced Photon Counting Techniques VIII},
editor = {Mark A. Itzler and Joe C. Campbell},
organization = {International Society for Optics and Photonics},
publisher = {SPIE},
pages = {911409},
year = {2014},
doi = {10.1117/12.2050808}
}

@article{pdc2003,
    author = {Kang, Y. and Lu, H. X. and Lo, Y.-H. and Bethune, D. S. and Risk, W. P.},
    title = {{Dark Count Probability and Quantum Efficiency of Avalanche Photodiodes for Single-Photon Detection}},
    journal = {Applied Physics Letters},
    volume = {83},
    number = {14},
    pages = {2955-2957},
    year = {2003},
    month = {10},
    doi = {10.1063/1.1616666}
}

@ARTICLE{spadpuf,
  author={Sajal, Md Sakibur and Dandin, Marc},
  journal={IEEE Transactions on Circuits and Systems I: Regular Papers}, 
  title={{PG-SPAD PUFs: Reconfigurable Physically Unclonable Functions Using Perimeter-Gated Single-Photon Avalanche Diodes}}, 
  year={2025},
  volume={},
  number={},
  pages={1-12},
  doi={10.1109/TCSI.2025.3607762}}

@article{Winsor1932TheCurve,
    title = {{The Gompertz Curve as a Growth Curve}},
    year = {1932},
    journal = {Proceedings of the National Academy of Sciences},
    author = {Winsor, Charles P.},
    number = {1},
    month = {1},
    pages = {1--8},
    volume = {18},
    doi = {10.1073/pnas.18.1.1},
    issn = {0027-8424}
}

@article{Dandin2010,
    title = {{Characterization of Single-Photon Avalanche Diodes in a 0.5 {$\mu$}m Standard CMOS Process—Part 1: Perimeter Breakdown Suppression}},
    year = {2010},
    journal = {IEEE Sensors Journal},
    author = {Dandin, Marc and Akturk, Akin and Nouri, Babak and Goldsman, Neil and Abshire, Pamela},
    number = {11},
    pages = {1682--1690},
    volume = {10},
    doi = {10.1109/JSEN.2010.2046163}
}

@article{Dandin2016,
    title = {{Characterization of Single-Photon Avalanche Diodes in a 0.5 {$\mu$}m Standard CMOS Process—Part 2: Equivalent Circuit Model and Geiger Mode Readout}},
    year = {2016},
    journal = {IEEE Sensors Journal},
    author = {Dandin, Marc and Habib, Mohammad Habib Ullah and Nouri, Babak and Abshire, Pamela and McFarlane, Nicole},
    number = {9},
    month = {5},
    pages = {3075--3083},
    volume = {16},
    doi = {10.1109/JSEN.2016.2526665}
}

@ARTICLE{sajalpbit,
  author={Sajal, Md Sakibur and Wu, Tong and Dandin, Marc and Srimani, Tathagata},
  journal={IEEE Electron Device Letters}, 
  title={{Probabilistic Bits Using Perimeter-Gated Single Photon Avalanche Diodes}}, 
  year={2025},
  volume={46},
  number={8},
  pages={1417-1420},
  doi={10.1109/LED.2025.3577944}}

@inproceedings{Sajal2022Perimeter-GatedProbability,
    title = {{Perimeter-Gated Single-Photon Avalanche Diode Imager with Vanishing Room Temperature Dark Count Probability}},
    year = {2022},
    booktitle = {2022 29th IEEE International Conference on Electronics, Circuits and Systems (ICECS)},
    author = {Sajal, Md. Sakibur and Lin, Kai-Chun and Senevirathna, Bathiya and Lu, Sheung and Dandin, Marc},
    pages = {1--4},
    publisher = {IEEE},
    doi = {10.1109/ICECS202256217.2022.9970824}
}

\end{document}